# Inverse-Designed Meta-Optics with Spectral-Spatial Engineered Response to Mimic Color Perception

*Chris Munley[1,+], Wenchao Ma[2,+], Johannes E. Fröch[1,3,+], Quentin A. A. Tanguy[3], Elyas Bayati[3], Karl F. Böhringer[3,5], Zin Lin[4], Raphaël Pestourie[4], Steven G. Johnson[3], Arka Majumdar[1,3]*

1: Department of Physics, University of Washington, Seattle, WA, USA

2: Department of Chemistry, Massachusetts Institute of Technology, Cambridge, MA, USA

3: Department of Electrical and Computer Engineering, University of Washington, Seattle, WA, USA

4: Department of Mathematics, Massachusetts Institute of Technology, Cambridge, MA, USA

5: Institute for Nano-engineered Systems, University of Washington, Seattle, WA, USA

+: Contributed Equally

**Abstract**

Meta-optics have rapidly become a major research field within the optics and photonics community, strongly driven by the seemingly limitless opportunities made possible by controlling optical wavefronts through interaction with arrays of sub-wavelength scatterers. As more and more modalities are explored, the design strategies to achieve desired functionalities become increasingly demanding, necessitating more advanced design techniques. Herein, the inverse-design approach is utilized to create a set of single-layer meta-optics that simultaneously focus light and shape the spectra of focused light without using any filters. Thus, both spatial and spectral properties of the meta-optics are optimized, resulting in spectra that mimic the color matching functions of the CIE 1931 XYZ color space, which links the distributions of wavelengths in light and the color perception of a human eye. Experimental demonstrations of these meta-optics show qualitative agreement with the theoretical predictions and help elucidate the focusing mechanism of these devices.

**Keywords:** dielectric metasurface, meta-optics, inverse design, color space, color sensor

1. Introduction

Over the recent years meta-optics have emerged as an elegant solution to control light in both near and far fields, enabling functionalities that would have been perceived as merely theoretically feasible decades back.[1] Their versatile capabilities are enabled by the arrangement of subwavelength scatterers, whose coordinated phase and amplitude response shape the properties of transmitted and/ or reflected light fields.[2–6] In the optical domain such sub-wavelength patterns have, for instance, been used to demonstrate lenses,[7–10] polarization control,[11–14] manipulation of light orbital angular momentum,[15–17] spectral filters,[18–20] and structural color prints.[21–23] To achieve these functionalities, forward design approaches are commonly utilized, where an analytical expression determines the arrangement of scatterers according to their local phase response (e.g. metalenses), or a scatterer geometry is optimized by a parameter sweep to achieve a wavelength dependent response (e.g. spectral filters). Yet, forward design approaches typically rely on intuition or analytical expressions, which often underlie certain boundary conditions. Moreover, as this research field progresses, requirements on meta-optic functionalities evolve to be progressively stringent as to become suitable for wide-scale practical applications.[24–26] Because with intuitive design approaches such multi-functional meta-optics are very difficult to realize, significant effort has been put forth in the field of computational nanophotonics to identify optimized geometries, leading to the concept of inverse design.[27] For meta-optics, this approach has recently been theoretically explored[28–34] and also used to experimentally demonstrate depth sensing by control of transmitted 3D optical fields,[35] extended depth of field metalenses in 1D[36] and 2D for full color imaging,[37] polychromatic large-area metalenses,[38] and for 3D printed meta-optics.[39] Most of these inverse designs are either used to tailor the *spatial* focal pattern of meta-optics at discrete wavelengths, or to provide a desired spectral response without any spatial mode engineering. However, for some applications, it would be desirable to simultaneously shape the *spectral* response and *spatial* response together, which is not yet reported using inverse design.

Here, we extend the inverse-design framework for meta-optics to simultaneously optimize spatial and spectral properties to design single-layer filter-free meta-optics that collect light in a way that mimics the color perception of a human eye. The functionality that we aim for is focusing a specific spectral distribution to a target area, namely, the color matching functions of the CIE 1931 XYZ color space,[40] which provides a link between spectra of light and colors perceived by the human eye. In more detail, the color perceived by a human eye is not solely a physical property of light but is determined by the responses of three types of cone cells in the retina of the eye. The combined feedback of these sensors enable the average human to perceive colors of light in the visible spectrum, which roughly ranges from 380 to 740 nm. Thus, to translate spectral measurements to human color experience, device-invariant representations of all colors visible to the average human eye are necessary, and as such the CIE color matching functions are often used to validate the color response of displays, assess the effect of pigments on the human eye, and represent the key to relating the objective and subjective natures of color.

Whereas the color perception of a human eye can be described by three color matching functions, we design three meta-optics, each of which corresponds to one of the color matching functions, as schematically depicted in Figure 1. These functions follow specific spectral distributions, which distinguish these meta-optics from traditional designs, such as metalenses with hyperbolic phase profiles. Specifically, the shape of the CIE 1931 X function can be approximated by a bimodal distribution, with local maxima near 450 nm and 600 nm, and an intensity ratio approximating 1:3.

The CIE 1931 Y and Z functions can be approximated with single modes centered near 550 nm and 450 nm, respectively.[41] However, these strongly differ by the width of the respective distributions, which are approximately 100 nm and 50 nm for the CIE Y and CIE Z function, respectively.

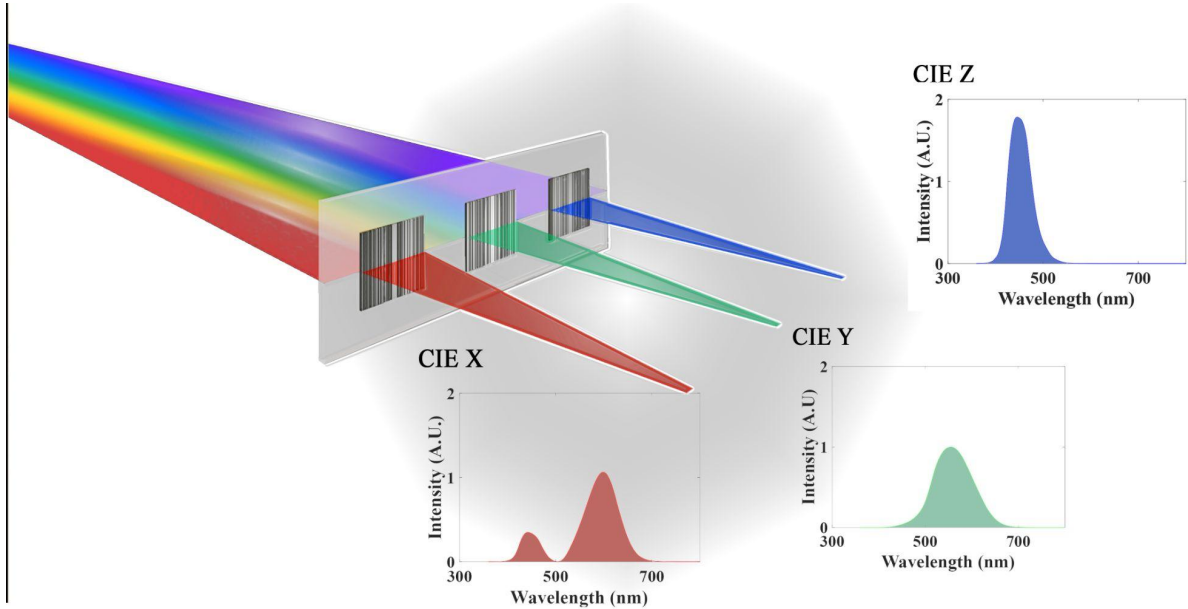

*Figure 1. Schematic of the CIE XYZ meta-optics. Three separate 1D meta-optics focus light with a spectral response equivalent to the CIE 1931 XYZ color space functions from left to right, respectively. The spectral responses are depicted for the respective color space functions.*

2. **Design**

To demonstrate the simultaneous optimization of spectral and spatial properties, we design 1D meta-optics as schematically shown in Figure 1. Each meta-optic consists of 3001 equally sized cells. Each cell, with a width of 361 nm and a length of 1 mm, contains a silicon nitride stripe on a silica substrate. The stripe width in each cell is used as a design parameter and allowed to vary between 100 nm and 261 nm. In detail, the far field $E_z^{far}(x, \lambda, w)$ in our target region can be computed as the convolution of the near field $E_z^{near}(x', \lambda, w)$ just above the meta-optic and the Green's function $G(x, x', \lambda)$, namely,

$$E_z^{far}(x, \lambda, w) = \int G(x, x', \lambda) E_z^{near}(x', \lambda, w) \, dx', \tag{1}$$

where $x'$ and $x$ denote spatial coordinates of the near- and far-field regions, with the direction parallel to the metasurface while perpendicular to the stripes; $\lambda$ denotes the wavelength; and $w = (w_1, w_2, \cdots, w_{3001})$ denotes the widths of the respective bars. Our aim is to optimize the widths $w$ such that the intensity $|E_z^{far}(x, \lambda, w)|^2$ in the target area $A$ approximates the color matching functions, which can be formulated as $\min_w$ objective($w$), where the objective function is

$$\text{objective}(w) = \left\| \frac{\int_A |E_z^{far}(x,\lambda,w)|^2 dx}{\int_A |E_z^{far}(x,\lambda_0,w_0)|^2 dx} - c\frac{f(\lambda)}{f(\lambda_0)} \right\|^2 = \int_{\lambda_{min}}^{\lambda_{max}} \left[ \frac{\int_{x_{min}}^{x_{max}} |E_z^{far}(x,\lambda,w)|^2 dx}{\int_{x_{min}}^{x_{max}} |E_z^{far}(x,\lambda_0,w_0)|^2 dx} - c\frac{f(\lambda)}{f(\lambda_0)} \right]^2 d\lambda. \tag{2}$$

Here, $\|\ \|$ denotes the 2-norm of a function of the wavelength $\lambda$ in $[\lambda_{min}, \lambda_{max}] = [380, 740]$ nm; the range of our specified far-field region is $[x_{min}, x_{max}] = [-12.5, 12.5]$ µm; $f(\lambda)$ denotes the analytical approximation of one of the color matching functions; $c$ is a dimensionless coefficient; $\lambda_0$ is the peak wavelength of the color matching function; and $w_0$ is the optimized widths that maximize the total intensity in $A$ at $\lambda_0$. The purpose of the scale factors $\int_A |E_z^{far}(x, \lambda_0, w_0)|^2 dx$ and $f(\lambda_0)$ is to nondimensionalize the objective function and to make the value of $c$ moderate. The value of $c$ is not a parameter under inverse design but needs to be determined optimizing $w$. Optimization with a large $c$ tends to result in high intensities at the specified far-field region but often leads to a poor match with the target function. On the other hand, a small $c$ results in lower intensities but a better match to the target function. The proper value of $c$, which yields a good match but with as large as possible intensities, can usually be determined via a few attempts (see Supplementary Information 1). Here, we use $c = 0.35$, $0.55$, and $0.7$ for CIE X, Y, and Z meta-optics, respectively.

Designing a large meta-optic with a brute-force solver is challenging. To reduce the computational cost, we apply a locally periodic approximation (LPA), under which $E_z^{near}$ of each unit cell is computed separately with periodic boundary conditions. We compute $E_z^{near}$ in a unit cell beforehand for 600 wavelengths at Gauss-Legendre quadrature points and 101 widths at Chebyshev nodes. At each wavelength, the dependence of $E_z^{near}$ on widths is fitted by Chebyshev interpolation. We thus obtain a surrogate model that quickly evaluates the near field above the meta-optic and hence facilitates the optimization.[28] The optimized intensity under LPA, the verification using finite-difference time-domain (FDTD) simulation, and the target functions are compared in Figure 2 a), b), and c) for the CIE X, Y, and Z meta-optics, respectively.

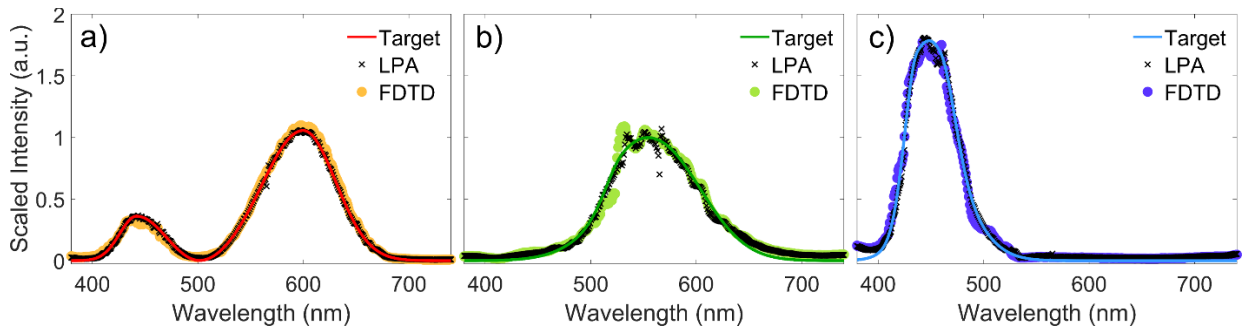

*Figure 2. Verification of the optimization based on LPA for the a) CIE X, b) CIE Y, and c) CIE Z meta-optic. The design results under LPA are the first terms in the brackets of Equation (2) multiplied by f(λ₀)/c.*

The optimized designs were then fabricated using a standard nanofabrication approach, in short consisting of deposition of 600 nm SiN on a fused quartz wafer, electron beam lithography, hard mask deposition, and reactive ion etching. Process parameters and conditions for individual steps are detailed in the methods section. Optical images of the fabricated devices are shown in Figure 3 a), b), and c) for the CIE X, Y, and Z meta-optics, respectively. Further details of the structures are revealed in scanning electron microscope (SEM) images at an oblique angle in

Figure 3 d), and e), which directly show the successful fabrication of features in the range of 100 to 261 nm for bars with lengths of 1 mm.

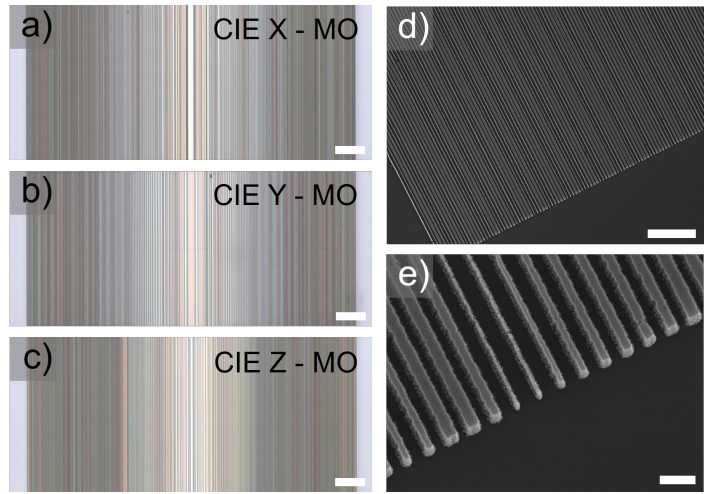

*Figure 3. Design and fabrication of the CIE meta-optics (MO). Optical images of the CIE X,Y, and Z meta-optics are shown in a), b), and c), respectively. The scale bars correspond to 100 μm. Magnified SEM images of the Y meta-optic at an oblique angle are shown in d) and e). The scale bars in d) and e) correspond to 5 μm and 500 nm, respectively.*

3. **Results and Discussion**

We now turn to the optical characterization of the fabricated devices. Collimated light from a stabilized halogen lamp (Thorlabs SLS302) transmitted through the sample substrate, then focused by the meta-optic, and was collected using an objective (Nikon Plan Fluor, 40X) with a numerical aperture of 0.75, which exceeds the numerical aperture of the meta-optics (0.3). The collected light was then guided and focused (Thorlabs AC254-30-A; focal length of 30 mm) onto a spectrometer (Princeton Instruments, IsoPlane-320; PIXIS 400B eXcelon), where an adjustable spectrometer slit served as a spatial aperture, limiting the collection area to a confined area in the focal plane.

The experimentally measured spectral dependence on the distance from the design focal plane (1.625 mm) is shown in Figure 4 a), b), and c) for the CIE X, Y, and Z meta-optics, respectively. For comparison, the corresponding simulation data are plotted in Figure 4 d), e), and f), where a focal area with the width of 25 µm is assumed. The experimental and simulation results show a good qualitative match in the spectral shifts with respect to the distance from the focal plane. The experimental results also exhibit a clear bimodal spectral distribution for the CIE X meta-optic and single modal distributions for the CIE Y and CIE Z meta-optics, which are characteristic of the target functions. In addition to these agreements, one can note that, in contrast to the simulations, the experimentally collected spectra show a cutoff around 425 nm, below which virtually no intensity was measured. This absence of light is likely related to the spectral distribution of the light source and the system response of the setup (See Supplementary Information 2). In the simulation results, one can also observe a demarcation, but around the wavelength of 523 nm, which is the width of a unit cell (361 nm) multiplied by the refractive index of the substrate (1.45). This feature related to diffraction is not present in the experimental data, probably because of the finite size of the meta-optics along the $y$ direction, which is assumed to be infinite in the simulation. Another apparent difference between simulation and experiment is a systematic redshift of approximately 25 nm to 50 nm, which may be related to fabrication imperfections, such as a mismatch of design and real refractive index, material thickness, or over/ under etching.

To demonstrate that the targeted functionality is achievable, we make a comparison of the desired CIE color matching functions, FDTD simulations, and spectra measured at a fixed distance (100 µm) from the design focal plane, as presented in Figure 4 g), h), and i) for the CIE X, Y, and Z meta-optics, respectively. The comparison shows a qualitative match between the target functions and the experimental results. We emphasize here that the inverse design proves successful as we simultaneously achieve focusing and spectral engineering, as evidenced by: a close spectral match in the proximity of the target focal plane for all three devices; distinct spectral features, such as the bimodal distribution of the transmitted light for the CIE X meta-optic with a node near 500 nm while maintaining the relative intensity distribution of the target function; and the characteristic spectral widths that, in all three cases, follow the trends of the target spectra. Particularly, the CIE Y meta-optic displays the broadest spectral distribution with a full width half maximum (FWHM) of ~ 100 nm, while a narrower spectrum was observed for the CIE Z meta-optic with a FWHM of ~ 50 nm, as intended.

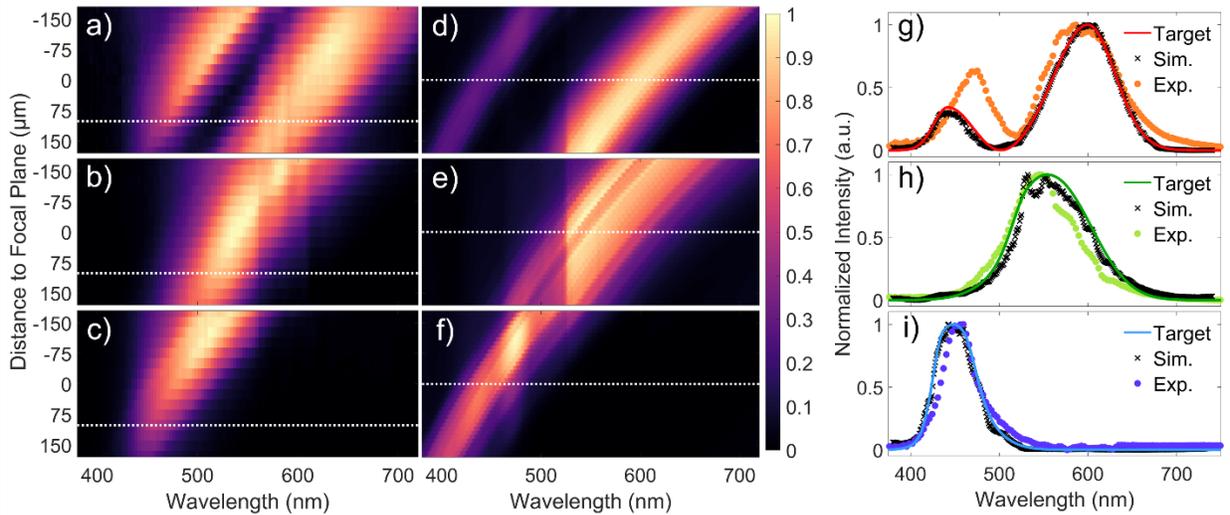

*Figure 4. Spectral response at the focal plane. The spectral response in the z direction is shown in a), b), and c) for the CIE X, Y, and Z meta-optics, respectively. Distance to focal plane indicates the distance to the design focal plane (1.625 mm). The white lines indicate the distance at which the spectra in g) - i) were collected. Simulated spectra in dependence of the distance to the design focal plane are shown in d), e), and f), for the CIE X, Y, and Z meta-optics, respectively. Experimental (exp.) spectra of the X, Y, and Z meta-optics are compared to simulation (sim.) and*

*to the analytical expression in g), h), and i), respectively. Target spectra are presented as full lines, simulated spectra are presented as x, and measured spectra are presented as dots.*

Because the inverse design process is non-intuitive, the emergence of complex spectral features requires further characterization. To unravel the underlying focusing mechanism we measure the lateral intensity profiles along the optical axis for transmitted light with defined spectral bands at (635 ± 6) nm, (530 ± 16) nm, and (455 ± 9) nm, from which we reconstruct a red (R), green (G), and blue (B) response (further detailed in the Methods).

The essential features of the intensity profiles are summarized in Figure 5 a) and b) for the R and B responses of the CIE X meta-optic, respectively, while Figure 5 c) shows the G response of the CIE Y meta-optic, and Figure 5 d) displays the B response of the CIE Z meta-optic, with the full intensity profiles presented in the Supplementary Information 4. For the R and B response of the CIE X meta-optic, we identify two major focal spots, which are separated by ~ 500 μm along the optical axis. The first focal spot of B is closely situated to the focal spot of R, while at the same time a lower field intensity can be assumed for wavelengths in between G and B, due to the separation along the optical axis of the wavelength dependent focal spots, which thus results in the bimodal shape of the target spectrum at the focal plane. Next, the CIE Y and Z meta-optics both possess one distinct focal spot, resulting in a spectral response with a unimodal distribution at the design focal plane. The main difference arises in the apparent focusing efficiency. The CIE Z meta-optic has a tighter focal spot for B in the focal plane as well as along the optical axis, whereas the focal spot of the CIE Y meta-optic is slightly extended in the focal plane and along the optical axis. As the focal spot is extended, the width of the spectral distribution also becomes extended, as the extended depth of field for the CIE Y meta-optic as compared to the CIE Z meta-optic leads to a relatively broader spectrum at the focal plane. Comprehensively, these distinct features underline the non-trivial focusing behavior of the CIE X, Y, and Z meta-optics, distinguishing them from traditional designs, such as the hyperbolic metalens.

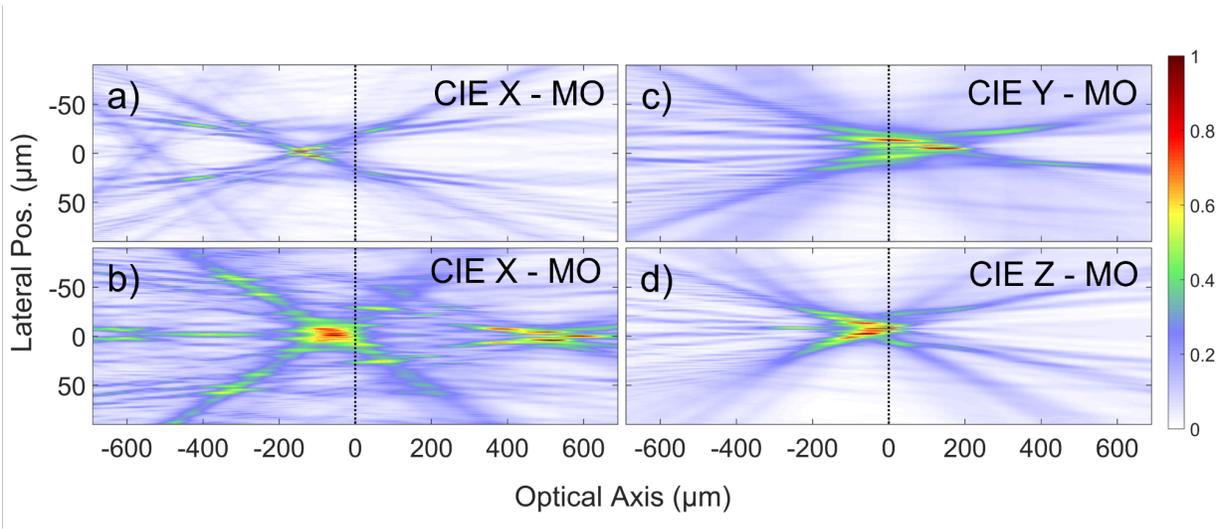

*Figure 5. Normalized intensity profiles for transmission of narrowband light sources plotted against the optical axis, relative to the design focal length (indicated by a dashed line). The intensity profiles correspond to a) red and b) blue transmission for the CIE X meta-optic, c) green transmission for the CIE Y meta-optic, and d) blue transmission for the CIE Z meta-optic, respectively.*

4. Conclusion

In summary, we have designed a set of single-layer filter-free 1D meta-optics with a 1 mm width. These meta-optics focus transmitted light to a specified area in the focal plane with a spectral response mimicking the CIE 1931 XYZ color matching functions. The measured spectral responses show qualitative agreement with the targets. Some key features of the target functions, such as the bimodal spectral distribution and the full widths at half maxima, are reproduced experimentally. We have also studied the focusing mechanism of the meta-optics from the spatial intensity profiles.

Through our work, we have demonstrated the capability to create devices that bridge the gap between the distributions of wavelengths in light and human color perception. We have thus combined the capabilities of focusing light with distinct broad spectral distributions that span the visible range. Beyond our work, these devices may be applicable in the future as elements to assist in devices for color sensing and in general pave the way for meta-optics with more complex functionalities.

**Methods**

**Simulations**

The framework of optimization, as reviewed above, is similar to those in References [28,34,36]. The near field in each unit cell is computed using rigorous coupled-wave analysis implemented in a free and open-source software package.[31] The data of the near field is then fitted to a Chebyshev-polynomial surrogate model. The convolution between the near field and the Green's function is performed by fast Fourier transforms.[42] The nonlinear optimization is performed by the CCSA-MMA algorithm[43] implemented in a free and open-source software package.[44] The final design is validated by finite-difference time-domain (FDTD) simulations also using a free and open-source software package.[45]

**Fabrication**

All devices were fabricated using the following fabrication process; first a 600 nm SiN film was deposited on quartz using plasma enhanced chemical vapor deposition (SPTS Delta LPX PECVD).

A positive polymer resist (ZEP 520) was spin coated, followed by deposition of a thin Au/Pd layer for charge dissipation. Electron beam lithography (JEOL JBX6300FS, 100 kV, 1 nA) was used to write the meta-optics layout in the resist. After exposure the Au/Pd layer was removed using TFA gold etchant and the resist was developed in Amyl Acetate for 2 minutes. Then an $AlO_x$ layer (~ 60 nm) was deposited in a thermal evaporator, followed by resist lift-off by ultrasonication in dichloromethane. The $AlO_x$ layer served as a hard mask in the following reactive ion etching step (Oxford Plasma Lab 100, ICP-180), where the SiN film was etched entirely. Finally, the devices were cleaned in a benign $O_2$ plasma.

**Optical Characterization**

For the presented spectra, the experimentally recorded spectra were corrected to account for background (i.e., light not diffracted by the meta-optic) and the spectral distribution of the light source (see Supplementary Information 2).

For measurements of the field profile in transmission the sample was mounted at a fixed position, and a microscope assembly consisting of an objective (Nikon Plan Fluor 20X, 0.5 NA), tube lens (Thorlabs, f=200 mm), and CMOS camera (Allied Vision ProSilica GT1930C) were aligned on a translation stage (NewPort ILS100CC). The position of the assembly was then swept in a range from 0 up to 3 mm with respect to the sample surface, while the lateral transmission intensity was captured in the objective focal plane for consecutive runs of the R, G, and B light sources.

**Acknowledgements**

Lorryn Wilhelm assisted with the design of Figure 1.

**Funding**

This research was supported by NSF-1825308, NSF-2127235, NSF-GCR-2120774, DARPA (contract no. 140D0420C0060), the U.S. Army Research Office through the Institute for Soldier Nanotechnologies at MIT under Award Nos. W911NF-18-2-0048 and W911NF-13-D-0001, and by the Simons Foundation collaboration on Extreme Wave Phenomena. RP is also supported in part by IBM Research via the MIT-IBM Watson AI Laboratory. Part of this work was conducted at the Washington Nanofabrication Facility / Molecular Analysis Facility, a National Nanotechnology Coordinated Infrastructure (NNCI) site at the University of Washington with partial support from the National Science Foundation via awards NNCI-1542101 and NNCI-2025489.

# Supplementary Information:

# Inverse-Designed Meta-Optics with Spectral-Spatial Engineered Response to Mimic Color Perception

Chris Munley[1,+], Wenchao Ma[2,+], Johannes E. Fröch[1,3,+], Quentin A. A. Tanguy[3], Elyas Bayati[3], Karl F. Böhringer[3,5], Zin Lin[4], Raphaël Pestourie[4], Steven G. Johnson[3], Arka Majumdar[1,3]

1: Department of Physics, University of Washington, Seattle, WA, USA

2: Department of Chemistry, Massachusetts Institute of Technology, Cambridge, MA, USA

3: Department of Electrical and Computer Engineering, University of Washington, Seattle, WA, USA

4: Department of Mathematics, Massachusetts Institute of Technology, Cambridge, MA, USA

5: Institute for Nano-engineered Systems, University of Washington, Seattle, WA, USA

+: Contributed Equally

**S1: Choice of values of *c***

Optimization with a large value for *c* tends to result in high intensities at the specified far-field region but often leads to a poor match with the target function, as shown in Figure S1 with *c* = 1. We choose the value of *c* that yields a good match but not too low intensity, as shown in Figure 2 in the main text. The variation of the minimized objective function and the maximum intensity with the value of *c* is illustrated in Figure S2.

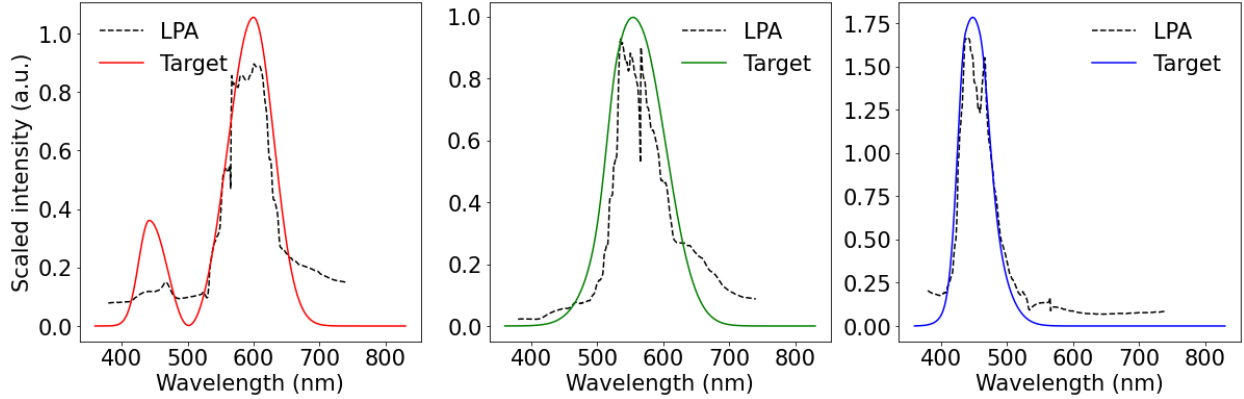

*Figure S1. Designed spectral patterns with c = 1. The three figures from left to right correspond to the CIE X, Y, and Z meta-optics.*

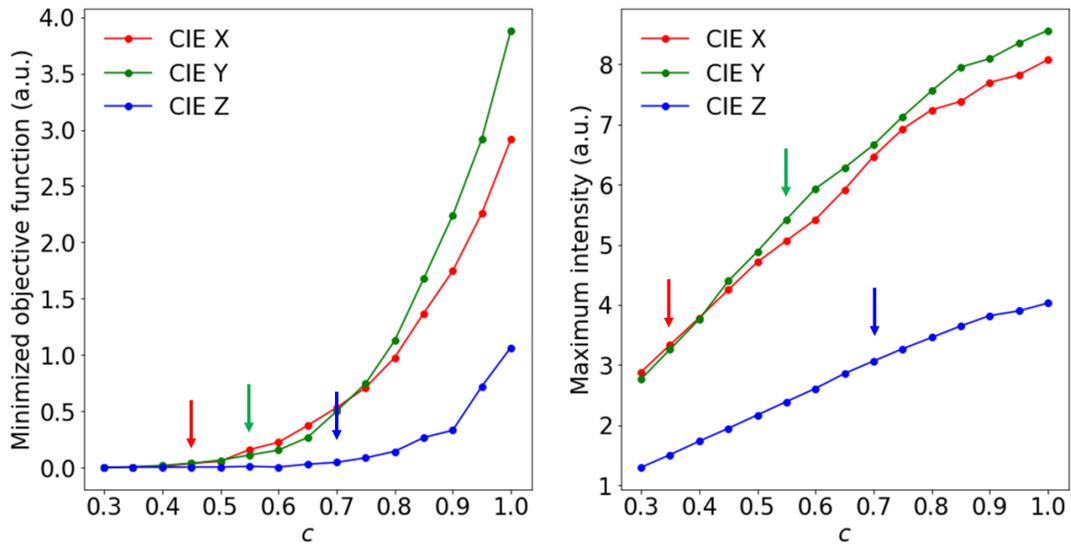

*Figure S2. Minimized values of objectives and maximum intensities as functions of the coefficient c. The chosen values of c, i.e., 0.35, 0.55, and 0.7 for CIE X, Y, Z, meta-optics, respectively, are indicated by arrows with the corresponding colors.*

## S2: Light Source Spectrum

To account for the spectral distribution of the light source (Thorlabs SLS302) and the wavelength dependent response of the optical setup, we measured the light source directly through the setup, as shown in Figure S3. Notably the intensity strongly decreases towards lower wavelengths and becomes virtually zero at wavelengths below ~ 420 nm, which explains the apparent cutoff observed in the experiment presented in Figure 4.

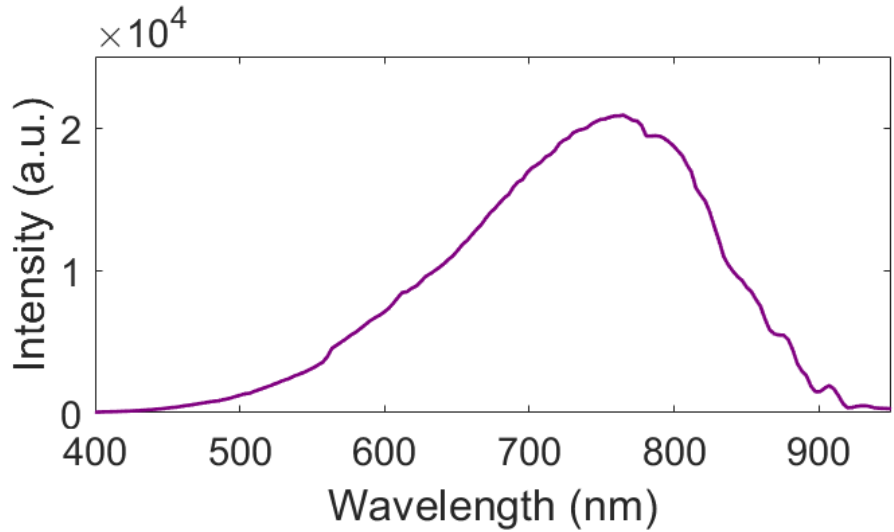

*Figure S3. Spectral distribution for the utilized light source as it was guided through the optical setup.*

**S3: Closest match to target function**

As described in the main text we observed good qualitative matches with the main spectral features of the target functions at a focal distance of around 100 μm. We furthermore observe even closer matches for the CIE XYZ meta-optics at slightly different distances relative to the focal plane. Specifically, as shown in Figure S4, we observe better matches at distances of ~152 um, 51 um, and 152 um, for the CIE X, Y, and Z meta-optics, respectively.

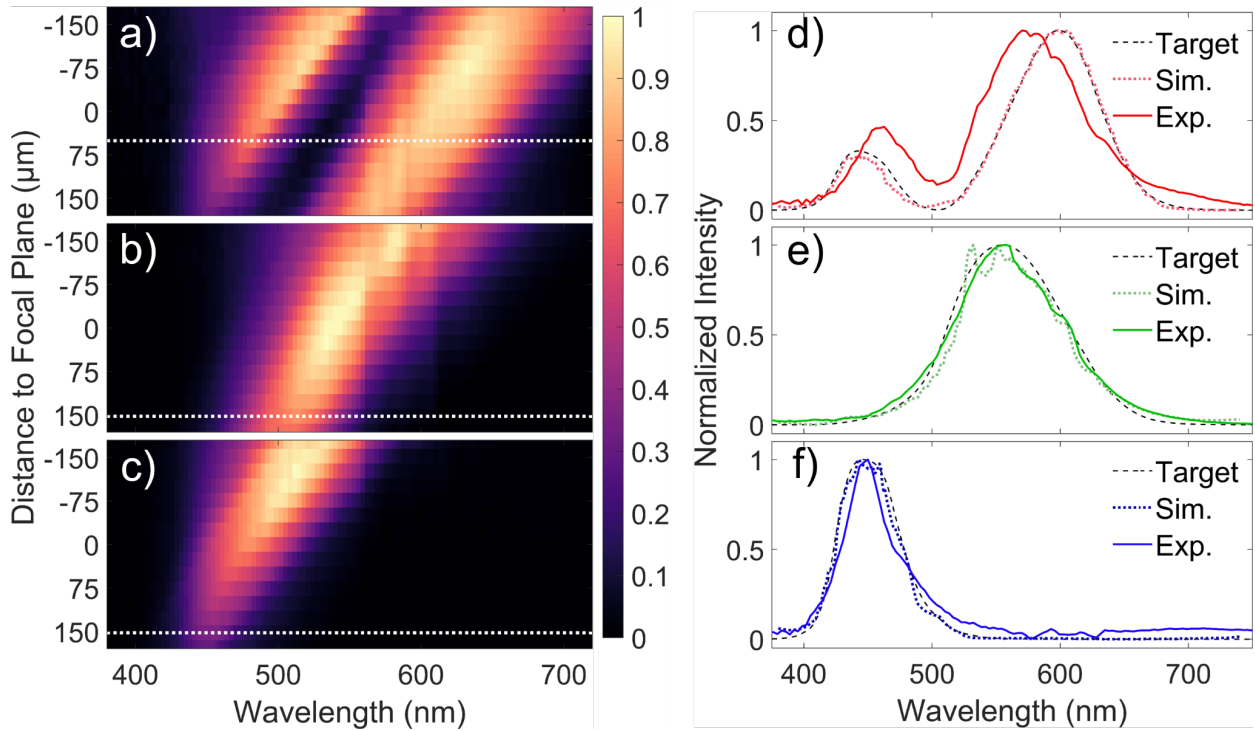

*Figure S4: Spectra of the closest match between target/ simulation and experiment for the CIE XYZ meta-optics. a), b), and c) show the spectral dependence with focal distance for the CIE X,Y, and Z meta-optics. The white dashed lines indicate the position of the closest match to the target function. d), e), and f) show the spectra of the closest matches in comparison to the target and simulation for the CIE X,Y, and Z meta-optics, respectively.*

## S4: Full RGB intensity profiles of the CIE XYZ meta-optics

The normalized transmission intensities are plotted with respect to the design focal distance (1.625 mm) in Figure S5, with the CIE X, Y, and Z meta-optics plotted column-wise and their respective response to R, G, and B, plotted row wise, respectively.

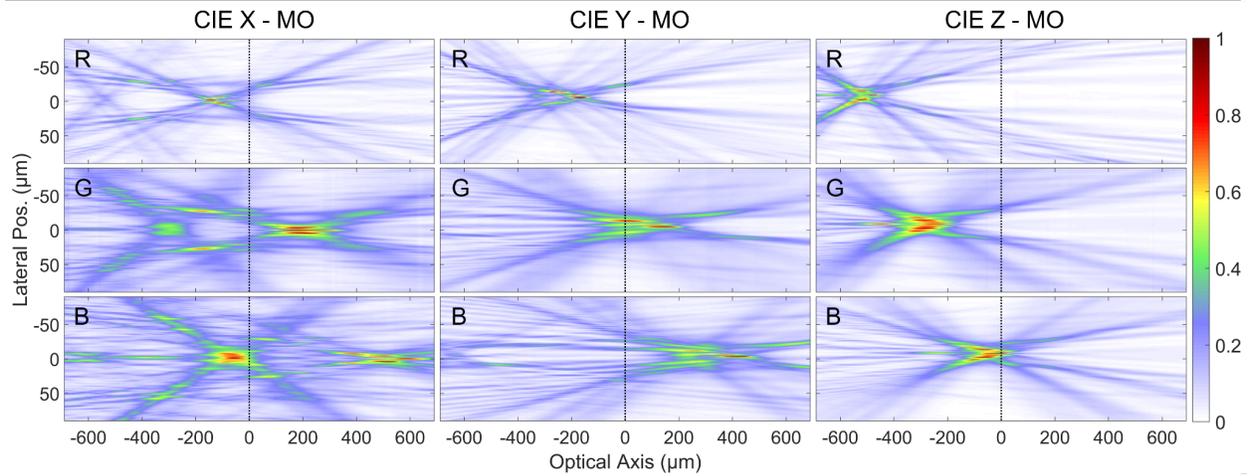

*Figure S5. Normalized intensity profiles for the transmission of red (R), green (G), and blue (B) light. The horizontal coordinate is the distance from the design focal plane (indicated by a dashed line) along the optical axis. The vertical coordinate is the distance from the focal axis along the lateral direction. Figures are presented row-wise for R, G, and B from top to bottom, and column-wise for the CIE X, Y and Z meta-optics from left to right.*